\documentclass[a4paper,reprint,aip,jcp]{revtex4-1}
\usepackage[T1]{fontenc}
\usepackage[utf8]{inputenc}

\usepackage{graphicx}
\usepackage[colorlinks,citecolor=blue,linkcolor=blue,urlcolor=blue]{hyperref}

\usepackage{bm}
\usepackage{siunitx}
\usepackage{chemmacros}
\usepackage{booktabs}

\begin{document}

\title{Resolving Dispersion and Induction Components for Polarisable Molecular Simulations of Ionic Liquids}

\author{Agílio A. H. Pádua}
\email{agilio.padua@uca.fr}
\affiliation{Institute of Chemistry of Clermont-Ferrand, Universté Clermont Auvergne \& CNRS, 63000 Clermont-Ferrand, France}

\date{\today}

\begin{abstract}
One important development in interaction potential models, or atomistic force fields, for molecular simulation is the inclusion of explicit polarisation, which represents the induction effects of charged or polar molecules on polarisable electron clouds. Polarisation can be included through fluctuating charges, induced multipoles or Drude dipoles. This work uses Drude dipoles and is focused on room-temperature ionic liquids, for which fixed-charge models predict too slow dynamics. The aim of this study is to devise a strategy to adapt existing non-polarisable force fields upon addition of polarisation, because induction was already contained to an extent, implicitly, due to parametrisation against empirical data. Therefore, a fraction of the van der Waals interaction energy should be subtracted so that the Lennard-Jones terms only account for dispersion and the Drude dipoles for induction. Symmetry-adapted perturbation theory (SAPT) is used to resolve the dispersion and induction terms in dimers and to calculate scaling factors to reduce the Lennard-Jones terms from the non-polarisable model. Simply adding Drude dipoles to an existing fixed-charge model already improves the prediction of transport properties, increasing diffusion coefficients and lowering the viscosity. Scaling down the Lennard-Jones terms leads to still faster dynamics and to densities that match experiment extremely well. The concept developed here improves the overall prediction of density and transport properties and can be adapted to other models and systems. In terms of microscopic structure of the ionic liquids, the inclusion of polarisation and the down-scaling of Lennard-Jones terms affects onyl slightly the ordering of the first shell of counterions, leading to small decreases in coordination numbers. Remarkably, the effect of polarisation is major beyond first neighbours, significantly weakening spatial correlations, a structural effect that is certainly related to the faster dynamics of polarisable models.
\end{abstract}

\maketitle

\section{Introduction}

The energetic, structural and dynamic proterties of condensed matter, and in particular fluid phases, can be described at different spatial and temporal scales.  Electronic structure calculations represent electrons explicitly using quantum mechanics, within the Born-Openheimer approximation, thus they include electronic and chemical effects. Atomistic force fields are classical models in which bonded and non-bonded interactions between atomic sites are represented by parametrised potentials. Parametrisation becomes necessary at scales for which it is unfeasible to compute interactions on the fly ab initio. At even larger scales, coarse-grain models are based on soft sites that correspond to entities larger than simple molecules or monomers. It is important to bridge these traditional scales in order to study phenomena that are not restricted to the usual theoretical levels. One such hybrid level is ``ab initio'' molecular dynamics using dispersion-corrected density functional theory (DFT), in which the van der Waals dispersion energy is added through classical London terms to the quantum DFT energies. Another is the incorporation of explicit polarisation (induction) in classical atomistic force fields, in an attempt to represent the distorsion of electron clouds induced by the electrostatic environment of the molecule or ion.

This is a topic in active development, notably though the use of Drude dipoles \cite{Lamoureux:2003wk} to represent polarisation of atomic sites. These methods have been available for some time in the CHARMM \cite{Brooks:2009ew} and NAMD \cite{Jiang:2011cm} molecular dynamics packages, intended for simulations using the Drude-2013 polarisable force field \cite{Lemkul:2016kv}. Recent implementations of Drude oscillators in Gromacs \cite{Lemkul:2015bf} and in LAMMPS \cite{Dequidt:2016kf} have become available, making polarisable force fields accessible in widely used molecular dynamics packages that allow simulation of biological molecules, organic compounds and many classes of materials. Other strategies to include explicit polarisation also in active development include charge equilibration methods \cite{Bauer:2012kd,Nistor:2006iq,Verstraelen:2009eg}.

Among general, fixed-charge atomistic force fields, OPLS-AA \cite{Jorgensen:1996bs} has been developed with a pragmatic philosophy, not only targeted at biological molecules but paying attention to the thermodynamic properties of simple liquids, exemplified by the Lennard-Jones parameters derived from experimental thermodynamic quantities such as densities and enthalpies of vaporisation. Other force fields, such as CHARMM/CGenFF \cite{Vanommeslaeghe:2010ch} or AMBER/GAFF, \cite{Cornell:1995bk,Wang:2004ue} have placed the emphasis on bio-macromolecules and less on liquid-state properties of simple molecules or ions.

For ionic liquids (ILs), one of the first general-purpose force fields \cite{CanongiaLopes:2004he,CanongiaLopes:2012ky} was developed based on the philosophy of OPLS-AA, ensuring transferability of parameters describing funcional groups between families of ions. This fixed-charge model has been successful in reproducing thermodynamic and solvation properties, but the predicted dynamics are too slow. In order to remedy this sluggish behaviour, several authors opted to scale the atomic partial charges by a factor of about 0.8 \cite{Bhargava:2007cm}. Another option is to add explicit polarisation to the force field, namely in the form of induced point dipoles \cite{Borodin:2009jb,Yan:2010jv} or Drude dipoles \cite{Schroder:2010jm}. Recent progress in this area includes detailed studies about how to attribute atomic polarisabilities in ionic liquids \cite{Gu:2013by,Bernardes:2015jz}, because those quantities are the basis for parametrising the Drude oscillators \cite{Lamoureux:2003wk}. The value of the point charges on an induced Drude dipole, $\pm q_D$, and the force constant of the harmonic bond between a Drude particle and its core atom, $k_D$, are related through the atom's polarisability, $\alpha$,
\begin{equation}
  \label{eq:pol}
  \alpha = \frac{q_D^2}{k_D}.
\end{equation}

Technically it is possible to add Drude dipoles to existing fixed-charged force fields, including OPLS-AA. In principle, the addition of Drude dipoles does not require changes in the partial electrostatic charges attributed to sites. However, adding polarisation should require that Lennard-Jones (or equivalent potential function) parameters be adapted, because in fixed-charge models the LJ potential accounts for repulsion, dispersion and polarisation, although implicitly for this last term. As such, simply adding Drude dipoles to a non-polarisable force field will lead to double counting the induction energies.

The aim of this work is to devise a strategy to evaluate the relative magnitude of dispersion and polarisation interactions in ionic liquids. We use symmetry-adapted perturbation theory (SAPT) \cite{Jeziorski:1994ef,Hohenstein:2011kd} to resolve the different components of the potential energy of interaction and propose a physically-sound method to scale-down the Lennard-Jones potential when polarisation is explicitly added to an existing force field. It is noteworthy that a recent polarisable force field for ionic liquids has been developed based on ab initio energies obtained using the the SAPT-DFT variant \cite{Choi:2014ba,McDaniel:2016gw}.

\section{Methods}

We adopted a fragment-based approach that allows us to span a variety of chemical structural units composing room-temperature ionic liquids. For example, we took 1-ethyl-3-methyl-imidazolium, \ch{C2C1im+}, as representative of cation head-groups in imidazolium ionic liquids; 1,1-dimethylpyrrolidinium, \ch{C1C1pyr+}, as cation head-group of pyrrolidinium ionic liquids; and butane as model compound for alkyl side chains that can be of different lengths. Anions, namely dicyanamide [\ch{N(CN)2-} or \ch{dca-}] and bis(trifluoromethanesulfonyl)amide (\ch{Ntf2-} or \ch{TFSI-}), were treated as entire molecular entities. These particular cations and anions were chosen because they compose common ionic liquids and because they lead to low-viscosity salts, which is beneficial for shorter simulation times.

The geometries of the isolated molecules or ions were first optimized using dispersion-corrected \cite{Grimme:2010ij} density functional theory (DFT) at the B97-D3/cc-pVDZ level. Then, dimers of cation-anion, cation-side chain or anion-side chain were optimised at the same level of theory.

The potential energy of interaction of each dimer was calculated at a series of distances, keeping the geometry of each monomer fixed. Since we intend to obtain relative quantities (induction with respect to dispertion energies) and not absolute ones, the main results should not be very sensitive to geometric details.  Interaction energies were calculated using a combination of SAPT method and basis set denoted sSAPT0/jaDZ, recommended as the ``bronze standard'' method for the calculation of interaction energies, and also with the higher level SAPT2+/aDZ combination, the ``silver standard'' method \cite{Parker:2014bg}. The sSAPT0 method describes intermonomer terms by perturbation to second order of dispersion (equivalent to MP2), leaving intramonomer terms at the Hartree-Fock level. The SAPT2+ method includes perturbation to second order in all intermonomer and intramonomer terms (describing intramonomer correlation at a level equivalent to MP4). The energies reported here for induction and dispersion include their exchange counterparts, as in eqs.~(7--9) of ref.~\onlinecite{Parker:2014bg}. Quantum calculations were performed using the Psi4 software \cite{Turney:2012gr}. Computational time with SAPT2+/aDZ is about two orders of magnitude (or worse) slower than with sSAPT0/jaDZ, with memory requirements also much higher, depending on the size of the fragments. As illustration, a single-point SAPT2+/aDZ calculation on the \ch{C2C1im+ ... Ntf2-} dimer took \SI{4}{days} on 16 processors requiring \SI{48}{GB} of memory, whereas the the same calculation at the sSAPT0/jaDZ took \SI{9}{min} and \SI{715}{MB}. Therefore, it is useful to assess the performance of the faster sSAPT0/jaDZ method to estimate the ratio of dispersion to induction energies, which is one quantity of interest here.

Molecular dynamics (MD) simulations of the ionic liquids \ch{[C2C1im][dca]}, \ch{[C4C1im][Ntf2]} and \ch{[C4C1pyr][Ntf2]} were performed using the LAMMPS package \cite{Plimpton:1995fc} with the CL\&P force field \cite{CanongiaLopes:2004he,CanongiaLopes:2012ky}, which is compatible with OPLS-AA \cite{Jorgensen:1996bs}. The parameters of the CL\&P force field are publicly available \cite{ilff:kz}. Starting configurations were prepared using the \texttt{fftool} \cite{fftool:kd} and \texttt{packmol} \cite{Martinez:2009di} utilities and consist of $N = 300$ ion pairs in periodic cubic boxes. The cutoff of pair interactions was \SI{12}{\angstrom}, with tail corrections applied, and electrostatic energies were evaluated using the PPPM method with an accuracy of $1 \times 10^{-4}$. The timestep was \SI{1}{fs}, bonds terminated in \ch{H} atoms were constrained using the SHAKE algorithm, and Nosé-Hoover thermostat and barostat were used to regulate temperature and pressure (always at \SI{1}{bar}).

Drude particles, including their interactions, neighbour lists and specific thermostat, where handled by the USER-DRUDE package \cite{Dequidt:2016kf} that implements thermalised, cold Drude dipoles in the LAMMPS code. Only ``heavy'' atoms (non \ch{H}) where treated as polarisable, with atomic polarisabilities taken from a recent study \cite{Bernardes:2015jz}. Thus, polarisabilities of \ch{H} atoms were summed into the heavy atoms to which they are bonded.  The mass displaced from the core atoms to the Drude particles was $m_D = \SI{0.4}{u}$ and the force constant of the bond between them was $k_D = \SI{4184}{kJ.mol^{-1}}$ (in the LAMMPS input file this value has to be halved), so the charges on Drude dipoles were derived from polarisabilities, with no adjustable parameters. The partial charges on the core atoms become the original atomic partial charges of the CL\&P force field minus the charges on the Drude particles. The interactions between neighbouring Drude oscillators were damped by Thole functions \cite{Dequidt:2016kf} with a coefficient $2.6$. The degrees of freedom corresponding to the distances between the Drude particles and their cores were thermostated at \SI{1}{K} \cite{Lamoureux:2003wk}. Keeping the Drude degrees of freedom at low temperature generates a trajectory close to that of relaxed (self-consistent) Drudes, enabling trajectories with a timestep of \SI{1}{fs} for the polarisable model, as was validated in the literature \cite{Lamoureux:2003wk,Dequidt:2016kf}.

Drude polarisation was added to the non-polarisable systems using the \texttt{polarizer} tool, distributed with the LAMMPS package and described in detail in the literature \cite{Dequidt:2016kf}. The interested reader can thus generate files with the non-polarisable and polarisable force field parameters used in the present work.

After initial \SI{0.5}{ns} equilibration runs, constant-$NpT$ trajectories of \SI{10}{ns} were generated for integer fixed-charge models (\SI{20}{ns} for \ch{[C4C1pyr][Ntf2]}), which have slower dynamics, and of \SI{5}{ns} for polarisable models. Simulation run times including Drude polarisation were a factor 3.5 slower than those of the respective non-polarisable systems (\SI{19}{h/ns} for polarisable against \SI{5.5}{h/ns} for fixed-charge simulations of \ch{[C4C1pyr][Ntf2]} on 16 processors using MPI parallelisation).

Diffusion coefficients were calculated from the mean-squared displacement using Einstein's relation,
\begin{equation}
  \label{eq:diff}
  D = \lim_{t\rightarrow\infty} \frac{1}{6}\frac{d}{dt} \left< (\bm{r}(t) - \bm{r}(0))^2 \right>,
\end{equation}
and viscosity from the components of the pressure tensor through the Green-Kubo relation,
\begin{equation}
  \label{eq:visc}
  \eta = \frac{V}{kT} \int_0^\infty \left< p_{xy}(t) p_{xy}(0) \right> dt.
\end{equation}
We averaged the viscosity values calculated over the off-diagonal components $p_{xy}$, $p_{yz}$ and $p_{xz}$.

\section{Results}

\begin{figure}[htb]
  \centering
  \includegraphics[width=8.5cm]{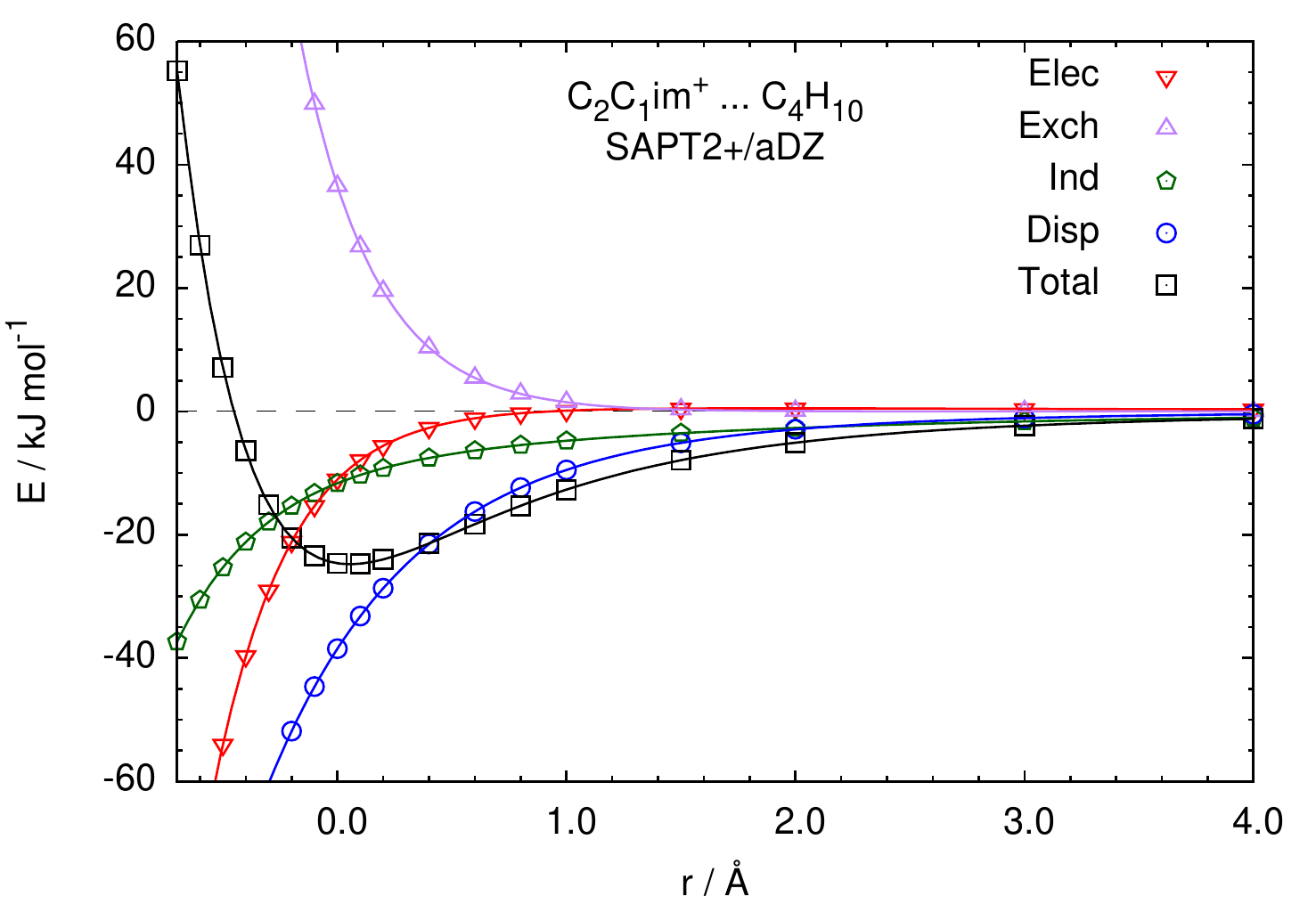}
  \includegraphics[width=8.5cm]{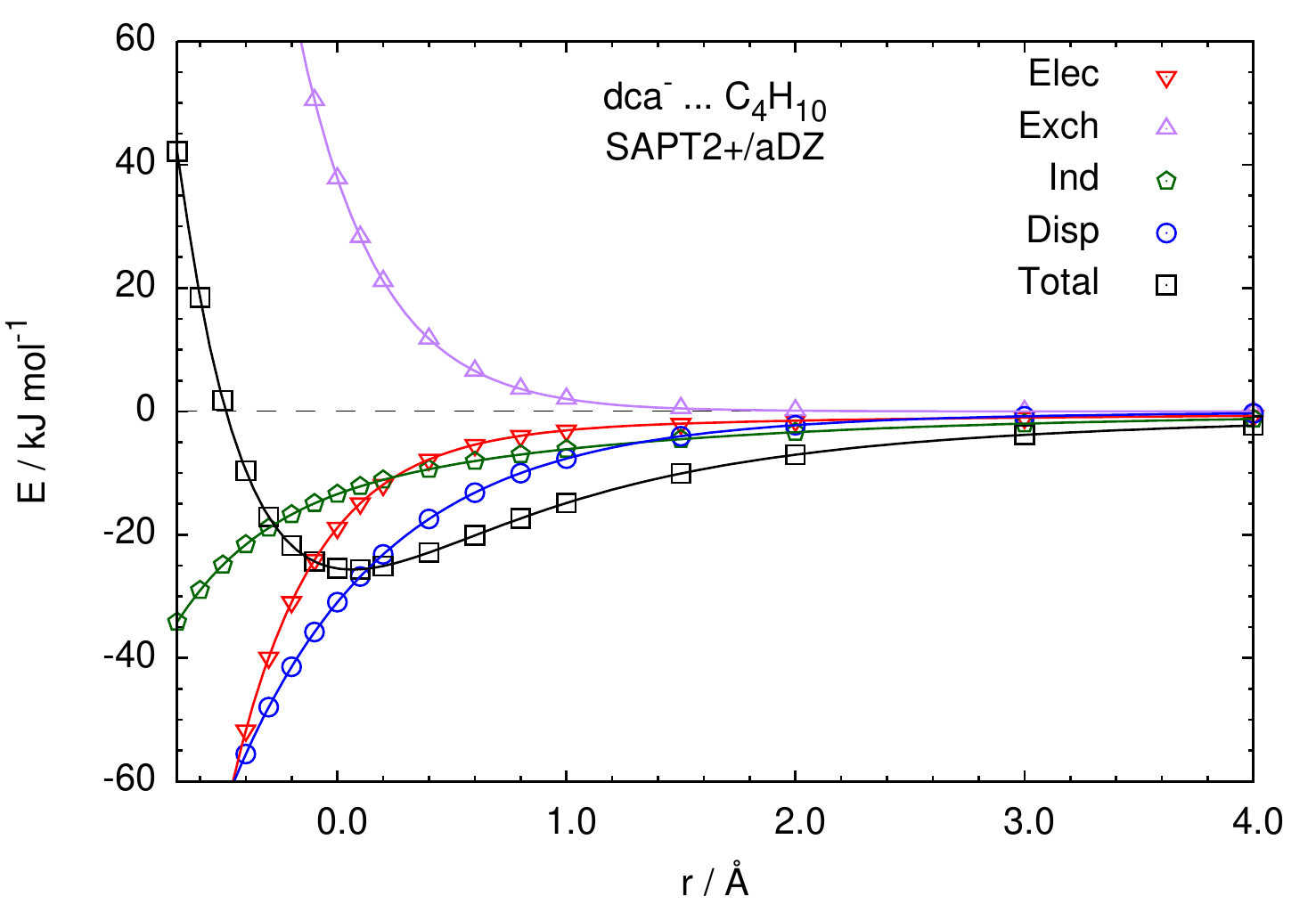}
  \includegraphics[width=8.5cm]{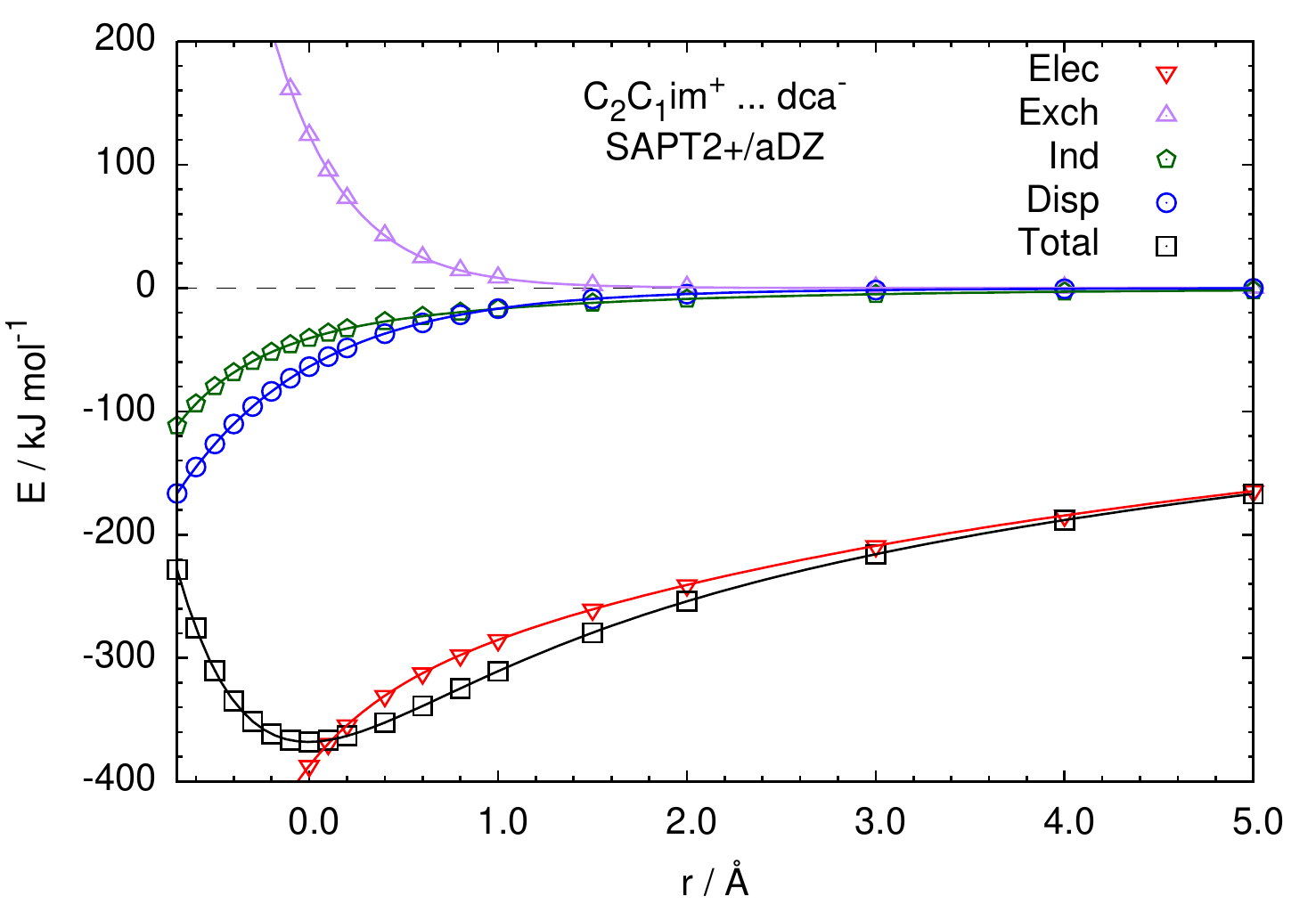}
  \caption{Decomposition of the potential energies of interaction involving dimers of \ch{C2C1im+}, \ch{dca-} and \ch{C4H10} from SAPT2+/aDZ. The distances are measured from the potential energy minima and the orientations were kept fixed from the optimised dimer. The lines are just guides to the eye.}
  \label{fig:sapt2} 
\end{figure}

The interaction potential curves presented in fig.~\ref{fig:sapt2} concern the \ch{C2C1im+} cation, the \ch{dca-} anion and \ch{C4H10} as a proxy for an alkyl side chain. These plots provide representative views of the energy decomposition into electrostatic, repulsive (exchange), induction (polarisation) and dispersion. The dominance of Coulomb terms in cation-anion interactions is evident in fig.~\ref{fig:sapt2}.

Whereas, qualitatively, the potential energy curves calculated at the sSAPT0/jaDZ level are similar to those obtained with SAPT2+/aDZ, the values of interaction energies obtained with the latter method are generally of higher magnitude (more negative). More important for the present study is the part of dispersion in non-Coulomb attraction (i.e. dispersion plus induction components), which we report in table~\ref{tab:dispind}.  We conclude that, even if in some cases close ratios are obtained using the two methods, there can be significant differences. Therefore, we think that the effective strategy is to use the fast sSAPT0/jaDZ in a series of calculations to find the distance corresponding to the potential energy minimum, and then use the more expensive SAPT2+/aDZ method only at the minimum geometry to evaluate the different components. In the dimers we studied, for which the SAPT2+/aDZ calculation was not too expensive, the distances of the minima with the two SAPT methods always coincided. The first entry in table~\ref{tab:dispind} contains the decomposition of interaction energy for the \ch{C4H10 ... C4H10} dimer, just to show that between non-polar fragments the attraction is essentially dispersive and the inductive part is very small, under \SI{1}{kJ.mol^{-1}}.

\begin{table}[htb]
  \centering
  \caption{Dispersion and induction energies of dimers obtained with different SAPT levels at the distance corresponding to the potential energy minimum. The factors $k_{ij}$ correspond to the fraction of dispersion in non-Coulomb attraction. Energies are in \si{kJ.mol^{-1}}.}
  \label{tab:dispind} 
  \begin{tabular}{lrrrr}
    \toprule
    Method & $E_\mathrm{tot}$ & $E_\mathrm{disp}$ & $E_\mathrm{ind}$ & $k_{ij}$ \\
    \midrule
    \multicolumn{5}{c}{\ch{C4H10 ... C4H10}} \\
    sSAPT0/jaDZ & $-5.49$ & $-9.29$ & $-0.60$ & $0.94$ \\
    SAPT2+/aDZ  & $-8.17$ & $-14.1$ & $-0.87$ & $0.94$ \\
    \midrule
    \multicolumn{5}{c}{\ch{C2C1im+ ... C4H10}} \\
    sSAPT0/jaDZ & $-19.8$ & $-27.6$ & $-9.16$ & $0.75$ \\
    SAPT2+/aDZ  & $-24.7$ & $-33.2$ & $-10.3$ & $0.76$ \\
    \midrule
    \multicolumn{5}{c}{\ch{C1C1pyr+ ... C4H10}} \\
    sSAPT0/jaDZ & $-13.1$ & $-17.3$ & $-9.35$ & $0.65$ \\
    SAPT2+/aDZ  & $-17.6$ & $-21.8$ & $-10.6$ & $0.67$ \\
    \midrule
    \multicolumn{5}{c}{\ch{dca- ... C4H10}} \\
    sSAPT0/jaDZ & $-18.0$ & $-17.2$ & $-9.58$ & $0.64$ \\
    SAPT2+/aDZ  & $-25.6$ & $-26.8$ & $-12.2$ & $0.69$ \\
    \midrule
    \multicolumn{5}{c}{\ch{Ntf2- ... C4H10} (O approaching)} \\
    sSAPT0/jaDZ & $-17.2$ & $-17.5$ & $-9.21$ & $0.66$ \\
    SAPT2+/aDZ  & $-25.0$ & $-26.1$ & $-10.5$ & $0.71$ \\
    \multicolumn{5}{c}{\ch{Ntf2- ... C4H10} (F approaching)} \\
    sSAPT0/jaDZ & $-8.31$ & $-8.53$ & $-2.52$ & $0.77$ \\
    SAPT2+/aDZ  & $-9.26$ & $-9.65$ & $-2.88$ & $0.77$ \\
    \midrule
    \multicolumn{5}{c}{\ch{C2C1im+ ... dca-}} \\
    sSAPT0/jaDZ & $-358.6$ & $-52.9$ & $-35.2$ & $0.60$ \\
    SAPT2+/aDZ  & $-368.0$ & $-63.8$ & $-40.8$ & $0.61$ \\
    \midrule
    \multicolumn{5}{c}{\ch{C2C1im+ ... Ntf2-}} \\
    sSAPT0/jaDZ & $-348.1$ & $-51.9$ & $-29.8$ & $0.64$ \\
    SAPT2+/aDZ  & $-362.1$ & $-72.3$ & $-38.5$ & $0.65$ \\
    \midrule
    \multicolumn{5}{c}{\ch{C1C1pyr+ ... Ntf2-}} \\
    sSAPT0/jaDZ & $-332.7$ & $-38.8$ & $-40.2$ & $0.49$ \\
    SAPT2+/aDZ  & $-342.9$ & $-55.7$ & $-48.1$ & $0.54$ \\
    \bottomrule
  \end{tabular}
\end{table}

We calculated factors $k_{ij} = E_\mathrm{disp} / (E_\mathrm{disp} + E_\mathrm{ind})$ that correspond to the fraction of dispersive attraction with respect to the sum of dispersion and induction, evaluated at the distance of minimum energy. Knowledge of the factors $k_{ij}$ provides a clue to scale down the Lennard-Jones well depths, so that this potential term will only account for dispersive attraction, leaving out the polarisation component to be represented explicitly by the Drude induced dipoles. Molecular dynamics simulations were performed for the ionic liquids \ch{[C2C1im][dca]}, \ch{[C4C1im][Ntf2]} and \ch{[C4C1pyr][Ntf2]} with three different force field settings: i) the original integer, fixed-charge CL\&P force field, ii) CL\&P with Drude oscillators added, iii) CL\&P with Drude oscillators and Lennard-Jones parameters $\epsilon_{ij}$ scaled down by the corresponding $k_{ij}$, according to the fragments involved. The choice to apply scaling factors only to interactions between cation and anion, or between a charged fragment and a side chain (neutral fragment) is justified because these interacting pairs can occur at short range, for which dispersion and induction are significant and differentiated. On the contrary, ions of the same charge are found most likely in their mutual second solvation shells due to charge ordering, therefore at distances beyond \SI{3}{\angstrom} in Fig.~\ref{fig:sapt2}, and there the short-range attractive terms become small.  Different temperatures were chosen for the MD simulations of the three ionic liquids because of the range of their viscosities. Since it would be difficult to obtain good statistics for the more viscous liquids at room temperature, the temperatures were adapted in the interest of reasonable simulation times and also considering the availability of experimental data for comparison.

Energetic quantities obtained by from the MD trajectories are listed in table~\ref{tab:eng}, and include the cohesive energy of the ionic liquids as well as the Lennard-Jones and electrostatic contributions. They were obtained by averaging over the MD trajectory of the liquids and deducing the intramolecular pair-wise additive contributions from the single ions (simulations of \SI{10}{ns} at constant $NVT$), thus for example $\left<E^c\right> = \left<E^\mathrm{IL}\right>/N - \left<E^+\right> - \left<E^-\right>$, where $N$ is the number of ion pairs in the condensed-phase simulation, and equivalently for $\left<E_\mathrm{LJ}\right>$ and $\left<E_\mathrm{elst}\right>$. Detailed values of the different energy terms are given in the Supplementary Information (Tab.~S1).

\begin{table}[htb]
  \centering
  \caption{Cohesive energy and component terms averaged over molecular dynamics trajectories in condensed phase. Electrostatic energy includes long-range part. The total cohesive energy $E^c$ includes intramolecular contributions. FixQ: integer fixed-charge CL\&P force field; Drude: polarisation added to the CL\&P model; SDrude: polarisation with scaled-down LJ $\epsilon_{ij}$.}
  \label{tab:eng}
  \begin{tabular}{lrrrrrc}
    \toprule
    ($E$/\si{kJ.mol^{-1}}) & $\left<E^c\right>$ & $\left<E_\mathrm{LJ}\right>$ &
    $\left<E_\mathrm{elst}\right>$ & $\left<E_\mathrm{self}\right>$ &
    $\left<E_\mathrm{ind}\right>$ & $K$ \\
    \midrule
    \multicolumn{7}{c}{\ch{[C2C1im][dca]} \SI{303}{K}} \\
    FixQ   & $-501.4$ &  $-72.2$ & $-429.8$ &   $0.0$ &   $0.0$ & $1.00$ \\
    Drude  & $-449.8$ &  $-67.1$ & $-444.8$ &  $27.1$ & $-42.1$ & $0.61$ \\
    SDrude & $-435.2$ &  $-45.8$ & $-454.9$ &  $30.6$ & $-55.7$ & $0.45$ \\
    \midrule
    \multicolumn{7}{c}{\ch{[C4C1im][Ntf2]} \SI{323}{K}} \\
    FixQ   & $-493.6$ & $-117.4$ & $-374.8$ &   $0.0$ &   $0.0$ & $1.00$ \\
    Drude  & $-423.0$ & $-112.3$ & $-364.8$ &  $39.7$ & $-29.7$ & $0.79$ \\
    SDrude & $-384.7$ &  $-81.8$ & $-366.9$ &  $41.3$ & $-33.4$ & $0.71$ \\
    \midrule
    \multicolumn{7}{c}{\ch{[C4C1pyr][Ntf2]} \SI{343}{K}} \\
    FixQ   & $-494.0$ & $-114.7$ & $-372.4$ &   $0.0$ &   $0.0$ & $1.00$ \\
    Drude  & $-407.4$ & $-105.8$ & $-351.9$ &  $37.1$ & $-16.6$ & $0.86$ \\
    SDrude & $-355.0$ &  $-60.6$ & $-355.2$ &  $38.9$ & $-21.7$ & $0.74$ \\
    \bottomrule
  \end{tabular}
\end{table}

Adding Drude polarisation to the fixed-charge model lowers the cohesive energy and leads to a slight reduction in the magnitude of Lennard-Jones (LJ) terms. When the LJ interactions are scaled down, the cohesive energy decreases further and an important decrease is observed in the LJ energy component in the simulated systems; at the same time the electrostatic part increases slightly. The induction term can be inferred, approximately, from the (negative) increase in electrostatic energy from the fixed-charge systems, taking also into account the self-energy of the Drude oscillators, $\left<E_\mathrm{ind}\right> \approx \left<E_\mathrm{elst}\right> - \left<E_\mathrm{elst}(\mathrm{FixQ})\right> - \left<E_\mathrm{self}\right>$. The term $\left<E_\mathrm{self}\right>$ is the energy required to create the induced dipoles, evaluated here through the potential energy stored in the harmonic springs between Drude particles and cores. So we are able to estimate the overall contribution of dispersion to the non-electrostatic (polarisation plus dispersion) energy in the condensed-phase systems. We adopt the same definition of the factor $k_{ij}$ as above, but now for the results of energy decomposition performed in condensed-phase configurations, averaged over the duration of the molecular dynamics trajectories. These resulting factors are denoted $K = \left<E_\mathrm{LJ}\right> / (\left<E_\mathrm{LJ}\right> + \left<E_\mathrm{ind}\right>)$ in table~\ref{tab:eng}.

Scaling down the LJ pair interactions in the polarisable model has a significant effect, leading to lower $K$ values. The consequences of the input $k_{ij}$ (affecting the pair interactions) on the condensed-phase energy decomposition are different in the three ionic liquids. In the \ch{dca-} IL an input $k_{ij} = 0.61$ affecting cation-anion pair interactions results in an overall $K = 0.45$ for the scaled-LJ model, so the down-scaling of LJ interactions is enhanced in the liquid-state energies (which have a stronger induction component). Conversely, in the \ch{Ntf2-} ILs, input $k_{ij} = 0.65$ or $0.54$ lead to $K = 0.71$ or $0.74$, respectively, so in these two ILs the induction component is smaller in the condensed phase than in the input interaction model. A complex picture was expected because the calculation of $K$ includes other interactions besides cation-anion pairs. But the behavior of \ch{[C2C1im][dca]} is opposite to our initial intuition: we thought that induction might be overestimated when calculated on an isolated dimer, in which interactions are more directional, than in a liquid phase where the local environment is more isotropic (as is expected also for charge transfer). Of course, the Drude oscillators were not parameterised based on the SAPT energies of this work; we are just analysing the link between the molecular interaction model and the resulting energy terms in the simulated liquid. Apart from this point, in all cases consistently, the induction energy in the liquids increases from the Drude model to the scaled-LJ Drude model, more significantly for \ch{[C2C1im][dca]} than for the two \ch{Ntf2-} ILs.

\begin{figure}[htb]
  \centering
  \includegraphics[width=8.5cm]{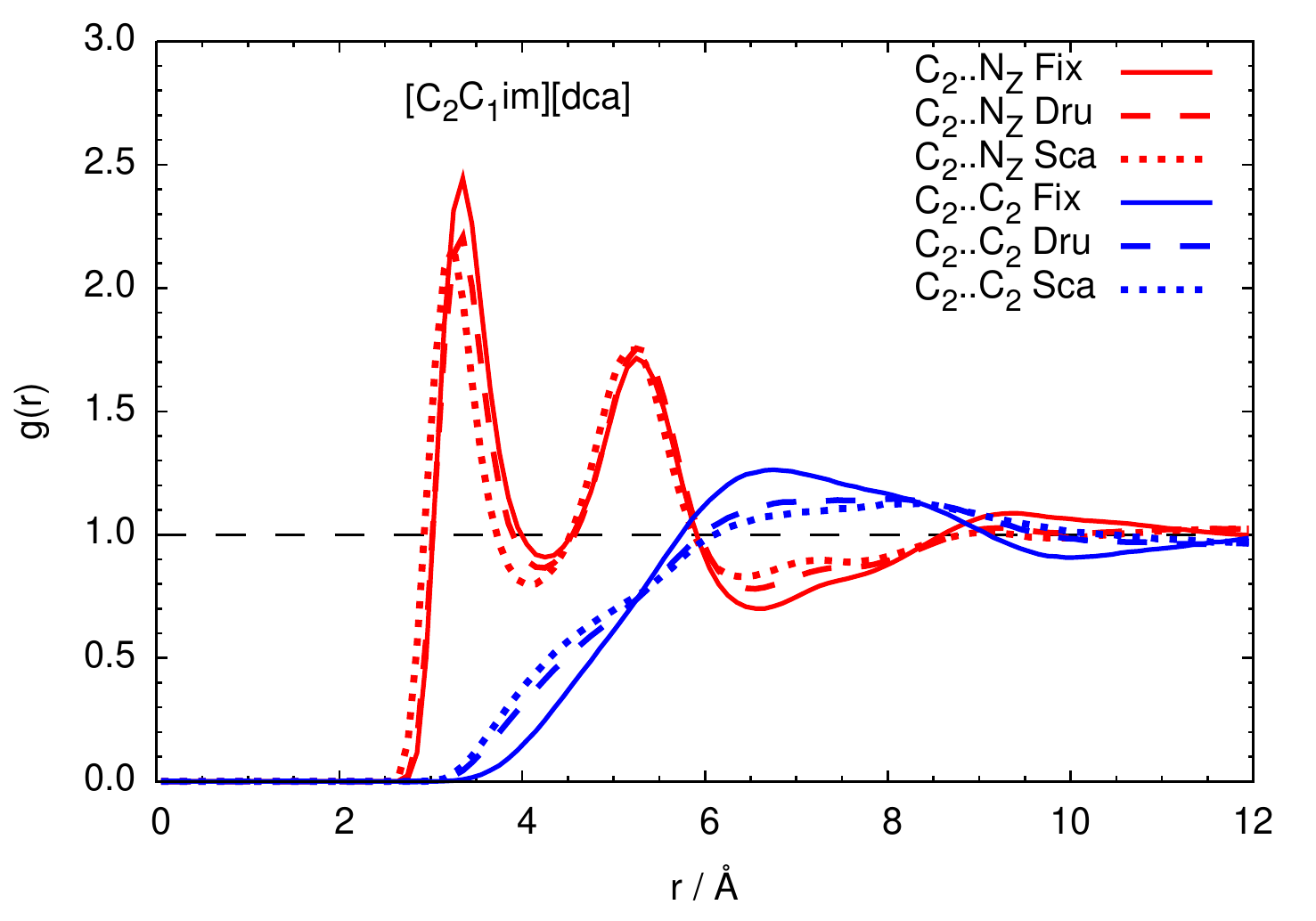}
  \includegraphics[width=8.5cm]{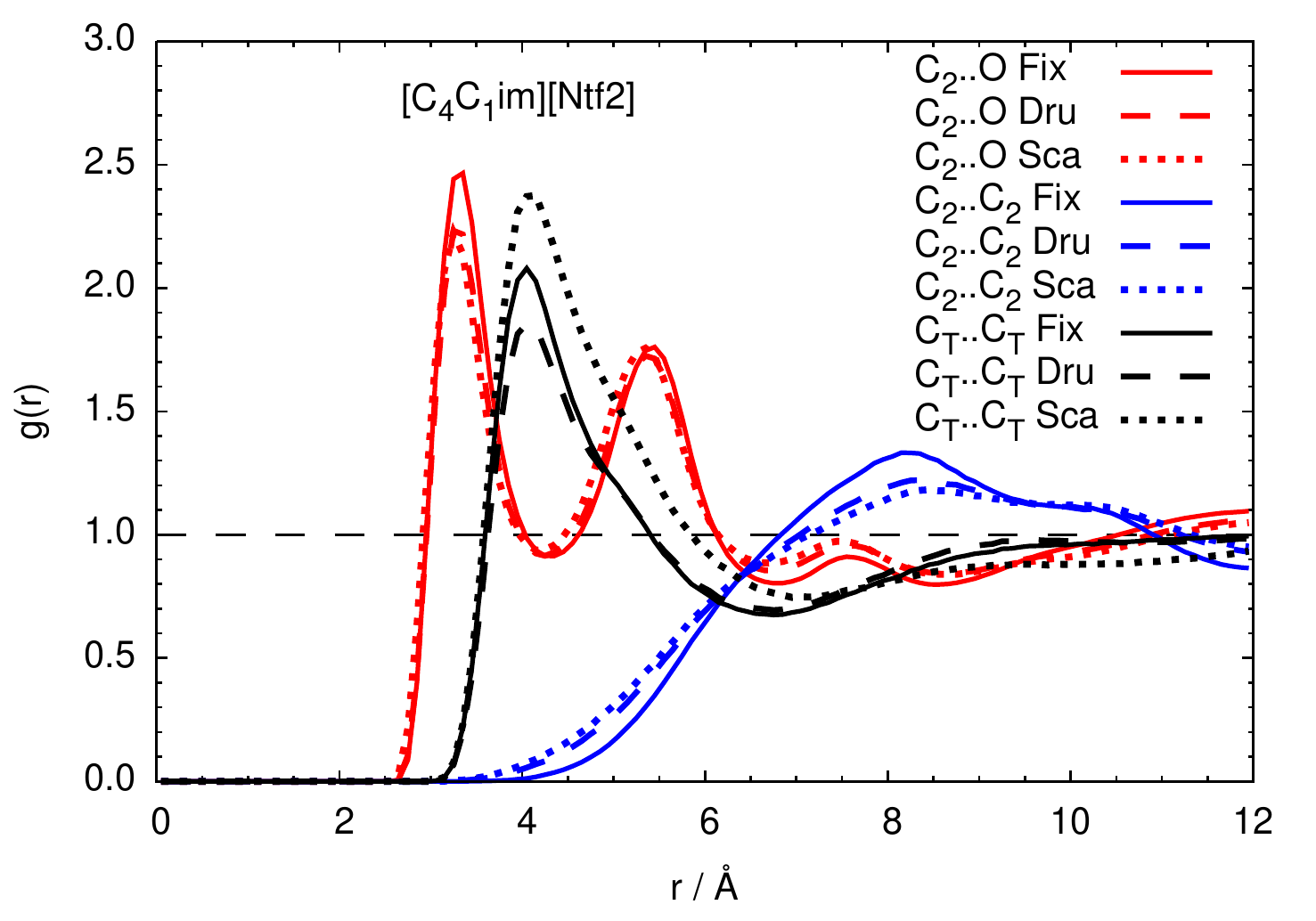}
  \includegraphics[width=8.5cm]{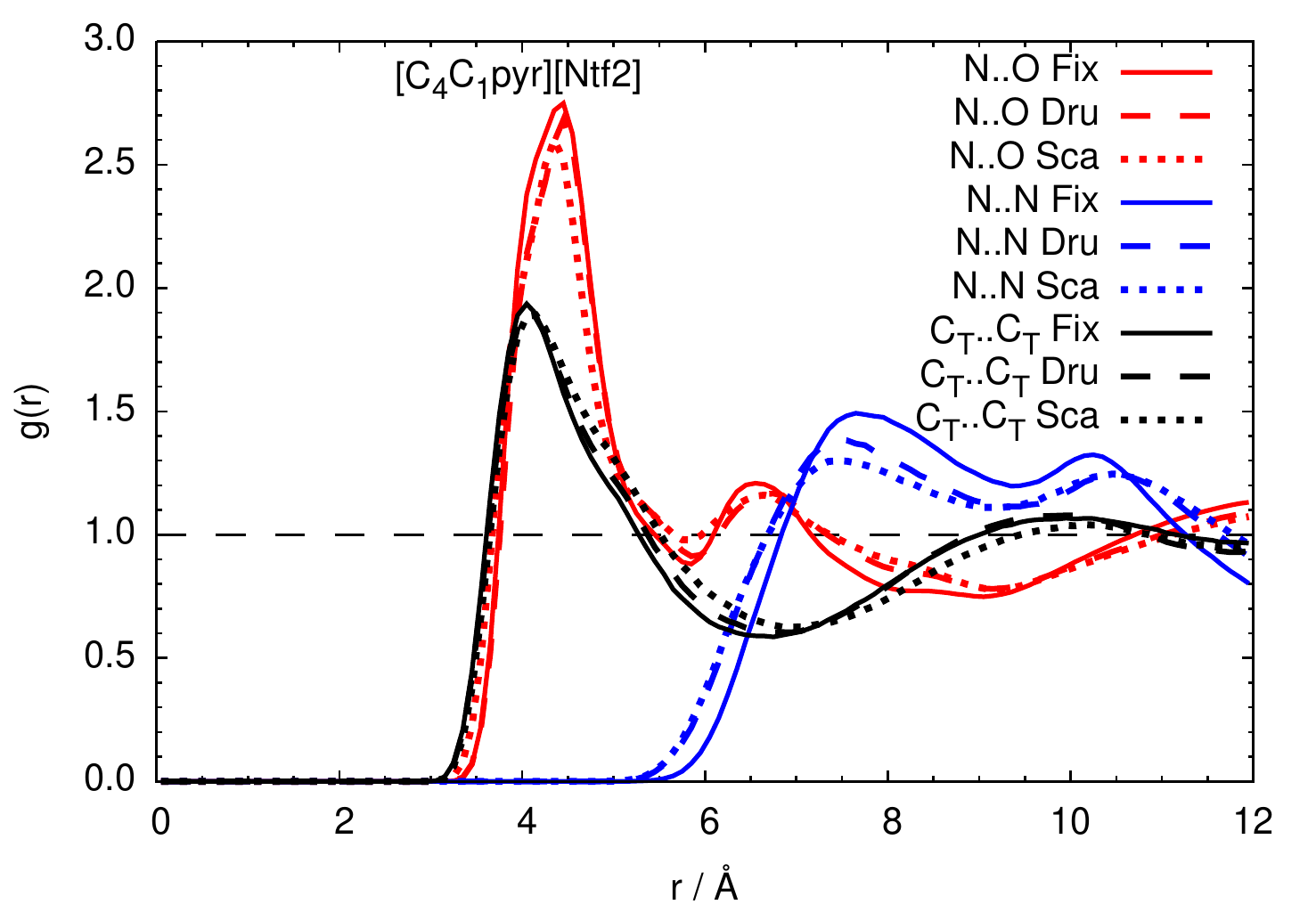}
  \caption{Radial distribution functions between representative atoms of the cation and the anion in \ch{[C2C1im][dca]} at \SI{303}{K}, \ch{[C4C1im][Ntf2]} at \SI{323}{K} and \ch{[C4C1pyr][Ntf2]} at \SI{343}{K}.}
  \label{fig:rdf}
\end{figure}

The effect of polarisation and of scaling the LJ potentials on the microscopic structure of the liquids is illustrated in figure~\ref{fig:rdf} by selected site-site radial distribution functions, $g(r)$, calculated in the three ionic liquids studied. We chose one positively-charged atom type representative of the imidazolium cation head-group (the \ch{C2} atom of the imidazolium ring), one negatively-charged atom type from the anions (the two terminal \ch{N_Z} atoms of \ch{dca-} and the four \ch{O} atoms of \ch{Ntf2-}), and the terminal atom of the butyl side chains (\ch{C_T}) to analyse eventual non-polar tail aggregation \cite{CanongiaLopes:2006ee}. These site-site pairs render the main features of local ordering of ions in the liquids.

\begin{figure}[htb]
  \centering
  \includegraphics[width=4.25cm]{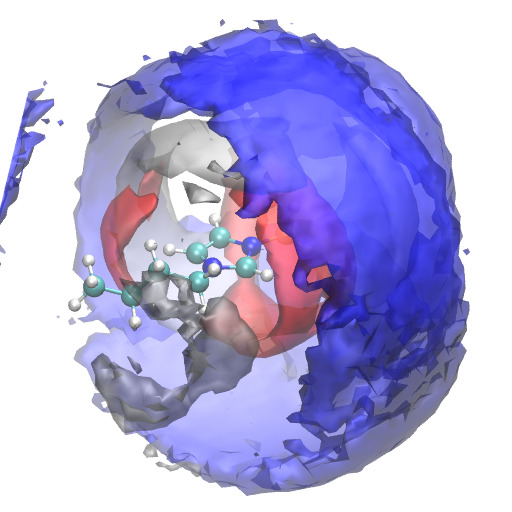}
  \includegraphics[width=4.25cm]{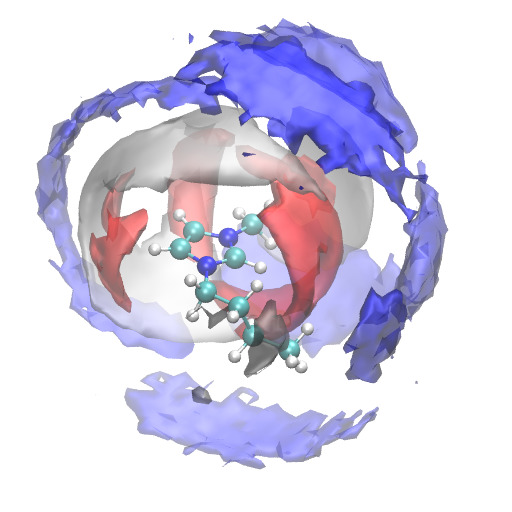}
  \includegraphics[width=4.25cm]{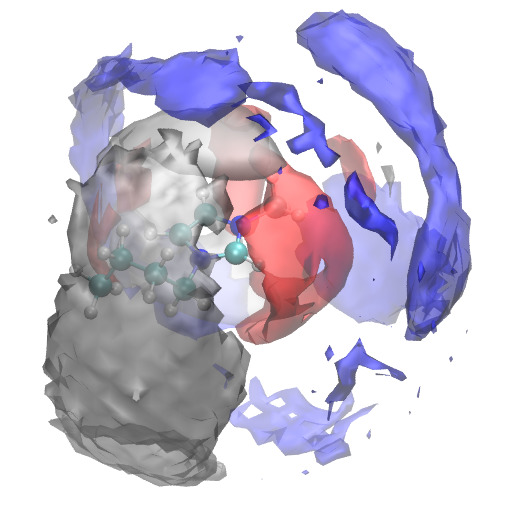}
  \caption{Spatial distribution functions of selected atoms around the cation in \ch{[C4C1im][Ntf2]} at \SI{323}{K}, respectively for FixQ; Drude; SDrude. The red surfaces correspond to \ch{O} atoms of the anions (isodensity contours at 4 times the average density). The blue surfaces correspond to \ch{C2} atoms of the imidazolium ring of cations (isodensity contours at 1.4 times the average density). The grey surfaces correspond to the terminal \ch{C} atoms of alkyl side chains of canions (isodensity contours at 1.5 times the average density).}
  \label{fig:sdf}
\end{figure}

\begin{table}[htb]
  \centering
  \caption{Coordination numbers of anion atoms up to first and second minima in the radial distribution functions around cation \ch{C2}. The site-site pairs considered are \ch{C2 ... N_Z} and \ch{C2 ... O}.}
  \label{tab:coord}
  \begin{tabular}{lcccrccccc}
    \toprule
    & \multicolumn{4}{c}{\ch{[C2C1im][dca]}} &
    &  \multicolumn{4}{c}{\ch{[C4C1im][Ntf2]}} \\
    ($r$/\AA) & $r_1$ & $n_{c1}$ & $r_2$ & $n_{c2}\ $ &
              & $r_1$ & $n_{c1}$ & $r_2$ & $n_{c2}$ \\
    \midrule
    FixQ   & $4.25$ & $2.44$ & $6.65$ & $10.09$ &
           & $4.25$ & $2.98$ & $6.75$ & $12.84$ \\
    Drude  & $4.25$ & $2.22$ & $6.55$ & $ 9.70$ & 
           & $4.25$ & $2.75$ & $6.75$ & $12.56$ \\
    SDrude & $4.05$ & $1.85$ & $6.45$ & $ 9.27$ &
           & $4.15$ & $2.51$ & $6.65$ & $11.98$ \\
    \bottomrule
  \end{tabular}
\end{table}

It is seen that most features of the cation-anion $g(r)$ remain similar among the three interaction models. If one takes the view of $g(r)$ centered on the cation \ch{C_2}, the two prominent peaks correspond to the two \ch{N_Z} atoms of \ch{dca-} or to the four \ch{O} atoms of \ch{Ntf2-}. The \SI{6.5}{\angstrom} distance after these two peaks encloses the first shell of counterions \cite{CanongiaLopes:2006ee}. The main difference due to the force fields is found in the intensity of the first peak, which decreases somewhat in the imidazolium ILs when polarisation is included (this effect is less noticeable in the pyrrolidinium IL). Also, the first peak occurs at slightly shorter distances with the polarisable models. Beyond this, the differences in peak height between the Drude model and the scaled-LJ polarisable model are small on cation-anion $g(r)$.  RDFs between the \ch{H2} atom of the cation and the \ch{N_Z} and \ch{O} atoms of the anions are shown in the suppelmentary information. In these $g(r)$ the first peak is more prominent, with small differences in intensity in \ch{[C2C1im][dca]} and almost none in \ch{[C4C1im][Ntf2]}. Again the polarisable models lead to slightly shorter distances. Coordination numbers up to the first and second minima are listed in table~\ref{tab:coord}. In \ch{[C2C1im][dca]} there are about 5 anions (10 \ch{N_Z}) surrounding each cation, whereas in \ch{[C4C1im][Ntf2]} 3 first-neighbour anions (12 \ch{O}) are found. Coordination numbers decrease slightly from fixed-charge to Drude models and also to the scaled-LJ model.

The more important and consistent differences in structure resulting from the different interaction models are seen not in first-neighbour cation-anion distributions, but in cation-cation second-shell spatial correlations. In figure~\ref{fig:rdf} the first peak in cation-cation $g(r)$ is more intense and localised in the fixed-charge model, and less intense and more spread-out (both towards shorter and longer distances) in the polarisable models, more so for the scaled-LJ model. The difference is significant in relative terms, i.e., compared to the peak intensity. In \ch{[C4C1im][Ntf2]} the $g(r)$ between terminal carbon atoms of the alkyl side chains indicates strong tail aggregation in the scaled-LJ model. No significant differences in tail aggregation resulting from the different interaction models are seen in \ch{[C4C1pyr][Ntf2]}.

Spatial distribution functions (SDFs) illustrate the three-dimensional structure of solvation layers. In figure~\ref{fig:sdf} we show distribution contour surfaces of selected atoms of anions or cations around central cations in \ch{[C4C1im][Ntf2]}. The distributions of \ch{O} atoms from anions around cations are quite similar but the spatial arrangement of cations around cations is very contrasted between the three interaction models, as expected from the RDFs. With the integer fixed-charge model, cations in second shell with respect to cations have an intense signal, which is reduced in the Drude model and further reduced (slightly) in the scaled-LJ model. In the scaled-LJ model the evidence of stronger alkyl chain aggregation is seen in the grey patch of terminal carbon atoms of other cations near the side chain of the central cations.

Therefore we see that although the structure of the first-neighbour shell is very similar between the three models, significant differences appear in the longer-range ordering. Including explicit polarisation through Drude oscillators and scaling down the LJ interactions leads to less localized spatial correlations between ions of the same charge, and the scaled-LJ model shows stronger side-chain aggregation.

Results for density, ion diffusion coefficients and viscosity are given in table~\ref{tab:md}, where they are compared with experimental data \cite{Franca:2014co,NietodeCastro:2010gr,Tiyapiboonchaiya:2003bi,Tokuda:2006gc,Castiglione:2009jy}. In terms of computational effort, polarisable simulations using Drude oscillators were for the present systems about 3.5 times slower than with the fixed-charge force field. However, this computational overhead is largely compensated because the sluggishness of dynamics is solved and shorter trajectories are required with the polarisable models to attain equivalent configurational sampling.

\begin{table}[htb]
  \centering
  \caption{Properties of ionic liquids obtained from molecular dynamics simulation using different versions of the force field, compared to experimental data.}
  \label{tab:md}
  \begin{tabular}{llcccc}
    \toprule
                          & $\rho$/\si{g.cm^{-3}} &
    $D_+$/\si{m^2.s^{-1}} & $D_-$/\si{m^2.s^{-1}} & $\eta$/\si{mPa.s} \\
    \midrule
    \multicolumn{5}{c}{\ch{[C2C1im][dca]} \SI{303}{K}} \\
    Exp    & $1.100$ & $1.4\times10^{-10}$ & $1.5\times10^{-10}$ & $13.9$ \\
    FixQ   & $1.103$ & $1.2\times10^{-11}$ & $1.4\times10^{-11}$ & $53\pm6$ \\
    Drude  & $1.085$ & $1.3\times10^{-10}$ & $1.4\times10^{-10}$ & $11\pm4$ \\
    SDrude & $1.086$ & $2.4\times10^{-10}$ & $2.3\times10^{-10}$ & $6.1\pm0.5$ \\
    \midrule
    \multicolumn{5}{c}{\ch{[C4C1im][Ntf2]} \SI{323}{K}} \\
    Exp    & $1.414$ & $6.6\times10^{-11}$ & $5.2\times10^{-11}$ & $20.6$ \\
    FixQ   & $1.484$ & $7.9\times10^{-12}$ & $6.1\times10^{-12}$ & $264\pm114$ \\
    Drude  & $1.469$ & $3.8\times10^{-11}$ & $2.7\times10^{-11}$ & $21\pm6$ \\
    SDrude & $1.425$ & $9.5\times10^{-11}$ & $6.9\times10^{-11}$ & $12\pm5$ \\
    \midrule
    \multicolumn{5}{c}{\ch{[C4C1pyr][Ntf2]} \SI{343}{K}} \\
    Exp   & $1.355$ & $10.2\times10^{-11}$ & $8.9\times10^{-11}$ & $16.2$ \\
    FixQ  & $1.417$ & $2.4\times10^{-12}$ & $2.1\times10^{-12}$ & $1052\pm875$ \\
    Drude  & $1.414$ & $1.2\times10^{-11}$ & $1.1\times10^{-11}$ & $526\pm372$ \\
    SDrude & $1.355$ &  $7.6\times10^{-11}$ & $7.5\times10^{-11}$ & $47\pm42$ \\
    \bottomrule
  \end{tabular}
\end{table}

When using the original CL\&P force field (FixQ) the density of liquid \ch{[C2C1im][dca]} obtained from simulation is very close to the experimental value, whereas the simulated densities of \ch{[C4C1im][Ntf2]} \ch{[C4C1pyr][Ntf2]} are higher than experiment by 4--5\%.  It is known that the CL\&P force field gives densities that are too high for some imidazolium bistriflamide ionic liquids \cite{CanongiaLopes:2004gt}, although when imidazolium or bistriflamide ions are combined with other counterions the densities agree better with experiment. (This was one case in which optimisation for a specific property and ionic liquid would have drawbacks in the transferability of the force field.) Adding Drude dipole polarisation leads to slightly lower densities for the three ionic liquids studied, but still not to agrement with experiment. Scaled-LJ with Drudes, however, leads to excellent agreement with experiment for the three ILs, with deviations within $\pm 1$\% in predicted densities.

Ion diffusion coefficients obtained with the CL\&P force field are much lower than the experimental values --- by one order of magnitude for \ch{[C2C1im][dca]} and \ch{[C4C1im][Ntf2]}, and even more for \ch{[C4C1pyr][Ntf2]} --- a well-known deficiency of non-polarisable force fields with integer ionic charges, which predict too slow dynamics.  Correspondingly, viscosities from simulation with fixed integer ionic charges are concomitantly higher than the experimental viscosities, as anticipated by a Stokes-Einstein argument that viscosity and diffusivity are inversely related. The fixed-charge model largely under-predicts diffusion and over-predicts viscosity for \ch{[C4C1pyr][Ntf2]}.

Adding Drude dipoles leads to faster diffusion, by one order of magnitude, bringing the calculated values into good agreement with experiment for \ch{[C2C1im][dca]} and \ch{[C4C1im][Ntf2]}, but still not for \ch{[C4C1pyr][Ntf2]}. The viscosities obtained with the Drude model are closer to the experimental values as well, although still far in \ch{[C4C1pyr][Ntf2]}, demonstrating the interest of explicit polarisation.  The scaled-LJ polarisable model produces even faster diffusion, when compared to added Drude dipoles without scaling of the LJ terms, overcorrecting a bit the diffusion coefficients, which become higher than the experimental ones in two cases. Viscosity is once more concomitantly reduced from the non-scaled result, in \ch{[C2C1im][dca]} and \ch{[C4C1im][Ntf2]} overcorrected such that the calculated values are below experiment, although the error bars almost encompass the experimental values. The best predictions for \ch{[C4C1pyr][Ntf2]} are with the scaled-LJ polarisable model, although the error bars are wider for this more viscous IL.

We conclude that adding explicit polarisation improves the prediction of transport properties (lower viscosity and faster diffusion) and that scaling-down the LJ terms leads to better densities and to even faster diffusion and lower viscosity. In \ch{[C2C1im][dca]} the scaled model gives dynamics that are a bit too fast, in \ch{[C4C1im][Ntf2]} diffusion is well predicted and viscosity a bit too low, and in \ch{[C4C1pyr][Ntf2]} diffusion is still a bit low and viscosity seems higher (although the large error bar prevents a clear conclusion). So, if we would increase the $k_{ij}$ values affecting the pair interactions, to give a larger weight to dispersion, we could improve the situation for \ch{[C2C1im][dca]} but would worsen it for \ch{[C4C1pyr][Ntf2]}, with respect to reproducing transport properties quantitatively. Scaling-down the LJ terms in the polarisable model by a justifiable factor, without additional tweaking, leads to an overall better agreement considering all the properties we calculate here, but further work on more systems and properties will be necessary to have a more complete view. The aim of this work was to devise and test a simple and physically-sound strategy, based on accessible quantum chemical calculations, to incorporate polarisation in existing atomistic force fields. This strategy can be adapted to other interaction models or systems.

The study of the effect of polarisation on the micsocopic ordering of the ionic liquids revealed unexpected features that are certainly linked to the reason why polarisation leads to faster dynamics. Although adding polarisation dos not change significantly the oerdring of the first shell of counteroins, the polarisable models showed significantly weaker spatial correlations beyond the first shell, namely in the cation-cation correlations.

\section{Supplementary Material}

Supplementary information to this manuscript contains additional plots of interaction potential energy between fragments of the ionic liquids obtained with the sSAPT0/jaDZ and SAPT2+/aDZ methods.

\begin{acknowledgments}
The author is thankful to Rahul Prasanna Misra (D. Blankschtein group at MIT), José Nuno Canongia Lopes (IST Lisbon), Christian Schröder (University of Vienna) and Alain Dequidt (Université Clermont Auvergne) for fruitful exchanges and suggestions.
\end{acknowledgments}

\bibliography{saptdrude}

\end{document}